\documentclass[12pt]{article}
\usepackage{amssymb}
\usepackage{amsmath}
\usepackage{amsfonts}
\usepackage{lmodern}						
\usepackage{latexsym}

\newcommand{\be}{\begin{equation}}
\newcommand{\ee}{\end{equation}}
\newcommand{\bn}{\begin{eqnarray}}
\newcommand{\en}{\end{eqnarray}}

\newcommand{\p}{\partial}
\newcommand{\dslash}{\partial\!\!\!/}

\newcommand{\al}{\alpha}
\newcommand{\ep}{\epsilon}
\newcommand{\ta}{\theta}
\newcommand{\de}{\delta}
\newcommand{\ga}{\gamma}

\def\no{\nonumber}

\def\[{\left\lbrack}
\def\]{\right\rbrack}

\def\({\left(}
\def\){\right)}
\def\ni{\noindent}	
\def\MyItem[#1]#2{\item[{#1}]#2}

\def\[{\left\lbrack}
\def\]{\right\rbrack}

\def\({\left(}
\def\){\right)}

\begin{document}

\begin{center}

\Large{\bf Some considerations on duality concerning kappa-Minkowski spacetime theories}

\bigskip 
\bigskip

Vahid Nikoofard$^{a,c}$\footnote{vahid@cbpf.br} and
Everton M. C. Abreu$^{b,c}$\footnote{evertonabreu@ufrrj.br}\\ \bigskip 

{\small
$^{a}$LAFEX, Centro Brasileiro de Pesquisas F\'{i}sicas, Rua Xavier Sigaud 150, 22290-180, \mbox{Rio de Janeiro}, RJ, Brazil\\
$^{b}$Grupo de F\'{i}sica Te\'orica e Matem\'atica F\'{i}sica, Departamento de F\'{i}sica, Universidade Federal Rural do Rio de Janeiro, 23890-971, Serop\'edica, RJ, Brazil\\
$^{c}$Departamento de F\'{i}sica, Universidade Federal de Juiz de Fora, 36036-330, Juiz de Fora, MG, Brazil\\}
\end{center}

\date{\today}

\begin{abstract}
\ni In this paper we have analyzed the $\kappa$-deformed Minkowski spacetime through the light of the interference phenomena in QFT where two opposite chiral fields are put together in the same multiplet and its consequences are discussed.  The chiral models analyzed here are the chiral Schwinger model, its generalized version and its gauge invariant version, where a Wess-Zumino term were added.  We will see that the final actions obtained here are in fact related to the original ones via duality transformations.
\end{abstract}

\newpage

\section{Introduction}

The fermion-boson mapping is one of the most investigated topics in theoretical physics during the last three decades due to its importance in the quantization of strings and also the Hall quantum effect. The possibility of mapping a complicated fermionic model into a scalar bosonic one was really attractive. This mapping is called bosonization and the chiral bosons can be obtained from the restriction of a scalar field to move in one direction only, as done by Siegel \cite{siegel}, or by a first-order Lagrangian theory, as proposed by Floreanini and Jackiw \cite{jackiw}. 

In two dimensions (2D), scalar fields can be viewed as bosonized versions of Dirac fermions and chiral bosons can
be seen to correspond to two-dimensional versions of Weyl fermions. As a generalization, in supergravity models, the extension of the chiral boson to higher dimensions has naturally introduced the concept of the chiral p-forms. 
Harada, in \cite{Harada}, has investigated the chiral Schwinger model via chiral bosonization and he has analyzed its spectrum. He has showed how to obtain a consistent coupling of FJ chiral bosons with a U(1) gauge field, starting from the chiral Schwinger model and discarding the right-handed degrees of freedom by means of a restriction in the phase space implemented by imposing the chiral constraint $\pi=\phi'$. 
In \cite{Bellucci}, Bellucci, Golterman and Petcher introduced an O(N) generalization of Siegel’s model for chiral bosons coupled to Abelian and non-Abelian gauge fields. The physical spectrum of the resulting Abelian theory is that of a (massless) chiral boson and a free massive scalar field. 
Initially, several models were suggested for chiral bosons but latter it was shown that there are some relations between these models \cite{Abreu:1998tc}. For instance, the Floreanini-Jackiw (FJ) model is the chiral dynamical sector of the more general model proposed by Siegel. The Siegel modes (rightons and leftons) carry not only chiral dynamics but also symmetry information. The symmetry content of the theory is described by the Siegel algebra, a truncate diffeomorphism, that disappears at the quantum level. As another application, chiral bosons appear in the analysis of quantum Hall effect \cite{ws}. The introduction of a soliton field as a charge-creating field obeying one additional equation of motion leads to a bosonization rule \cite{Girotti:1988ua}.

The direct sum of two chiral fermions in 2D gives rise to a full Dirac fermion, however this is not true for their bosonized versions as noticed in \cite{ms}, see also \cite{adw}. Besides, the fermionic determinant of a Dirac fermion interacting with a vector gauge field in D = 1 + 1 factorizes into the product of two chiral determinants but the full bosonic effective action is not the direct sum of the naive chiral effective actions as discussed in \cite{abw}. Stated differently, the action of a bosonized Dirac fermion is not simply the sum of the actions of two bosonized Weyl fermions, or chiral bosons. Physically, this is connected with the necessity to abandon the separated right and left symmetries,
and accept that vector gauge symmetry should be preserved at all times. This restriction will force the two independent chiral bosons to belong to the same \textit{multiplet}, effectively soldering them together. In both cases it turns out that an interference term between the opposite chiral bosonic actions is needed to achieve the expected result, such term is provided by the so called soldering procedure. 

The concept of soldering has proved extremely useful in different contexts \cite{w2}. This formalism essentially combines two distinct Lagrangians carrying dual aspects of some symmetry to yield a new Lagrangian which is exposed of, or rather hides, that symmetry. These so-called quantum interference effects, whether constructive or destructive, among the dual aspects of symmetry, are thereby captured through this mechanism \cite{abw}. The formalism introduced by M. Stone \cite{ms} could actually be interpreted as a new method of dynamical mass generation through the result obtained in \cite{abw}. This is possible by considering the interference of right and left gauged FJ chiral bosons. The result of the chiral interference shows the presence of a massive vectorial mode for the special case where the Bose symmetry fixed the Jackiw-Rajaraman regularization parameter as $a = 1$ \cite{jr}, which is the value where the chiral theories have only one massless excitation in their spectra. This clearly shows that the massive vector mode results from the interference between two massless modes.

It was shown lately \cite{aw}, that in the soldering process of two opposite chiral fields, a lefton and a righton, coupled to a gauge field, the gauge field decouples from the physical field. The final action describes a non-mover field (a noton) at the classical level. The noton acquires dynamics upon quantization. This field was introduced by Hull \cite{hull} to cancel out the Siegel anomaly. It carries a representation of the full diffeomorphism group, while its chiral components carry the representation of the chiral diffeomorphism.

The same procedure works in $D = 2 + 1$ if we substitute chirality by helicity. For instance, by fusing together two topologically massive modes generated by the bosonization of two massive Thirring models with opposite mass signatures in the long wave-length limit. The bosonized modes, which are described by self and anti-self dual Chern-Simons models \cite{tpn,dj}, were then soldered into two massive modes of the $3D$ Proca model \cite{bw}. More generally, the $\pm 1$ helicity modes may have different masses which leads after soldering to a Maxwell-Chern-Simons-Proca (MCSP) theory. In this case, technical problems \cite{bk} regarding a full off-shell soldering can be resolved by defining a generalized soldering procedure \cite{gs}.

The basic idea of the generalized soldering is the introduction of a free parameter $\alpha$ with a sign freedom which plays a role whenever interactions are present. In the soldering of two chiral Schwinger models that results either to an axial $(\alpha = −1)$ or to a vector $(\alpha = +1)$ Schwinger model which are dual do each other. In the case of the two Maxwell-Chern-Simons theories, the choice of the $\alpha $-parameter with opposite sign leads to dual interaction terms. We can have either a derivative coupling or a minimal coupling plus a Thirring term. After integration over the soldering field the dependence on the sign of $\alpha$ disappears which proves that they correspond to dual forms of the same interacting theory.  Recently, a new idea concerning the construction of the so-called Noether vector, the concept of can be directly analyzed from an initial master action \cite{clovis}.   We will discuss this issue here in the future. 

Recently the soldering formalism was used to investigate the self-dual theories with spin $s \geq 2$ and opposite helicities. In \cite{Dalmazi:2009es, Dalmazi:2010bf} the authors have demonstrated that the linearized Fierz-Pauli action which describes a doublet of massive spin-2 particles can be obtained via a soldering procedure of two second order self-dual models of opposite helicities. Besides, one can recover the New Massive Gravity NMG \cite{Alishahiha:2014dma, Bergshoeff:2009hq} (also at the linearized level) by soldering two self-dual models of opposite helicities of either third or fourth order in derivatives.

Usually the noncommutativity can be implemented by using the Weyl operators or, for the
sake of practical applications,  through the way of normal functions with a suitable definition of
star-products \cite{Douglas:2001ba}. Generally the noncommutativity of spacetime may be encoded
through ordinary products in the NC $^\star$-algebra of Weyl operators, or equivalently
through the deformation of the product of the commutative C$^\star$-algebra of functions
to a NC star-product. For instance, in the canonical NC  spacetime
the star-product is simply the Moyal-product \cite{moyal}, while on the $\kappa$-deformed Minkowski
spacetime the star-product requires a more complicated formula \cite{AmelinoCamelia:2001fd}. 

In order to treat the $\kappa$-deformed Minkowski spacetime in a very similar way to the usual Minkowski spacetime, the authors in \cite{Miao:2009fu, Miao:2011um} have proposed a quite different approach to the implementation of noncommutativity. To this aim a well-defined proper time  from the $\kappa$-deformed Minkowski spacetime has been defined that corresponds to the standard basis. In this way we encode enough information of noncommutativity of the $\kappa$-Minkowski spacetime to a commutative spacetime in this new parameter, and then set up a NC extension of the Minkowski spacetime. This extended Minkowski spacetime is as commutative as the Minkowski spacetime, but it contains noncommutativity already. Therefore, one can somehow investigate the NC field theories defined on the $\kappa$-deformed Minkowski spacetime by following the way of the ordinary (commutative) field theories on the NC extension of the Minkowski spacetime, and thus depict the noncommutativity within the framework of this commutative spacetime. With this simplified treatment of the noncommutativity of the $\kappa$-Minkowski spacetime, we unveil the fuzziness in the temporal dimension and build noncommutative chiral boson models in \cite{Miao:2009fu}.

The organization of the issues through this paper obeys the following sequence: in section 2 we have written a review of the $\kappa$-Minkowski noncommutativity and in section 3, a review of the essentials of the soldering formalism.  In section 4, we have analyzed the soldering of the NC chiral 
Scwinger model (CSM) and in section 5, the NC version of the generalized CSM.  In section 6 we have discussed the gauge invariant CSM.  The (anti)self-dual model in $D=2+1$ was analyzed in section 7.  As usual, the conclusions and perspectives were described in the last section.


\section{The NC extension of Minkowski spacetime}

The commutative spacetime is characterized by the canonical Heisenberg commutation relations
\begin{equation}
\left[ \hat{\mathcal{X}}^\mu,\hat{\mathcal{X}}^\nu\right]=0 , \hspace{.7cm} \left[ \hat{\mathcal{X}}^\mu,\hat{\mathcal{P}}_\nu\right]=i\delta_\nu^\mu, \hspace{.7cm}  \left[ \hat{\mathcal{P}}_\mu,\hat{\mathcal{P}}_\nu\right]=0 \label{commut}
\end{equation}  

\ni where $\mu,\nu=0,1,2,3$. In order to introduce the $\kappa$-deformed Minkowski spacetime we have \cite{Miao:2009fu}
\begin{eqnarray}
\hat{x}^0=\hat{\mathcal{X}}^0-\frac{1}{k}\left[ \hat{\mathcal{X}}^i,\hat{\mathcal{P}}_j\right]_+, \hspace{.8cm} \hat{x}^i=\hat{\mathcal{X}}^i+A \: \eta^{ij}\hat{\mathcal{P}}_j \: exp(\frac{2}{k}\hat{\mathcal{P}}_0) 
\end{eqnarray} 
where $\left[\hat{\mathcal{O}}_1, \hat{\mathcal{O}}_2\right]_+\equiv \frac{1}{2}(\hat{\mathcal{O}}_1\hat{\mathcal{O}}_2+ \hat{\mathcal{O}}_2\hat{\mathcal{O}}_1)$,  $\eta^{\mu\nu}\equiv diag(1,-1,-1,-1)$, $i,j=1,2,3$ and $A$ is an arbitrary constant. The NC parameter $\kappa$ has mass dimension and it is real and positive. The Casimir operator related to the $\kappa$-deformed Poincaré's algebra is
\begin{equation}
\hat{\mathcal{C}}_1=\left( 2k sinh\frac{\hat{p}_0}{2k}\right) ^2-\hat{p}^2_i, \label{casimir}
\end{equation} 
and for the momentum operators we have
\begin{equation}
\hat{p}_0=2k \: sinh^{-1}\frac{\hat{\mathcal{P}}_0}{2k}, \hspace{.9cm} \hat{p}_i=\hat{\mathcal{P}}_i.
\end{equation}
With  these last results we can construct our NC phase-space $\left(\hat{x}^{\mu},\hat{p}_{\nu} \right) $
\begin{eqnarray}
\left[\hat{x}^0,\hat{x}^j \right]=\frac{i}{k}\hat{x}^j , \hspace{.7cm} \left[\hat{x}^i,\hat{x}^j \right]=0,\hspace{.7cm} \left[\hat{p}_{\mu},\hat{p}_{\nu} \right]=0, \hspace{.7cm} \left[\hat{x}^i,\hat{p}_j \right]=i\delta^i_j \hfill \\
\left[\hat{x}^0,\hat{p}_0 \right]=i \left( cosh\frac{\hat{p}_0}{2k} \right)^{-1}, \hspace{.7cm} \left[\hat{x}^0,\hat{p}_i \right]=-\frac{i}{k}\hat{p}_i, \hspace{.7cm} \left[\hat{x}^i,\hat{p}_0 \right]=0
\end{eqnarray}
which satisfies the Jacobi identity. It is easy to see that when $k\rightarrow\infty$ we recover the commutative phase-space in Eq.(\ref{commut}).

The Casimir operator described above in Eq.(\ref{casimir}) can now be written in the standard way
\begin{equation}
\hat{\mathcal{C}}_1=\hat{\mathcal{P}}_0^2-\hat{\mathcal{P}}_i^2
\end{equation}

\ni where it is easy to see that this selection coincides with the ones in Eq.(\ref{commut}).   In the case that $\hat{p}_{\mu}$ has standard forms like
\begin{equation}
\hat{p}_{0}=-i\frac{\partial}{\partial t}, \hspace{.7cm} \hat{p}_{i}=-i\frac{\partial}{\partial x^i},
\end{equation}
so that the operator $\hat{\mathcal{P}}_{0}$ then reads
\be
\hat{\mathcal{P}}_{0}=-2ik\left(sin\frac{1}{2k}\frac{\partial}{\partial t} \right). \label{nc-mom0}
\ee
In \cite{Miao:2009fu} the author has introduced a proper time $\tau$ through the operator 
\be
\hat{\mathcal{P}}_0\equiv -i\frac{\partial}{\partial \tau} \label{nc-mom}
\ee
and using Eqs. (\ref{nc-mom0}) and (\ref{nc-mom}) we have that 
\be
2k\left(sin\frac{1}{2k}\frac{d}{dt} \right)\tau=1 
\ee

\ni which solution is
\be \label{trans}
\tau=t+\sum \limits_{n=0}^{+\infty}c_{-n} \: exp(-2kn\pi t)
\ee
where $n\geq 0$, $n\in \mathbb{N}$. The coefficients $c_{-n}$ are arbitrary real constants. This property implies a kind of temporal fuzziness coherent in the $\kappa$-Minkowski spacetime. Notice that as $k\rightarrow \infty$, the proper time turns back to the ordinary time variable.

To construct a NC extension of Minkowski spacetime $(\tau, x^i)$(where the NC feature is inside the proper time), let us define a twisted $t$-coordinate, such that the metric is
\bn \label{metric}
g_{00}&=&\dot{\tau}^2=\left[ 1-2k\pi\sum \limits_{n=0}^{+\infty}nc_{-n} \: exp(-2kn\pi t)\right]^2 \nonumber \hfill \\
g_{11}&=&g_{22}=g_{33}=-1.
\en
So, we can use Eq.(\ref{metric}), construct NC models in the commutative framework. Namely, we construct a Lagrangian theory for NC model in the extended framework of the Minkowski spacetime.

\section{The canonical soldering formalism}\label{section5}

The basic idea of the soldering procedure is to raise a global Noether symmetry of the self and anti-self dual constituents into a local one, but for an effective composite system, consisting of the dual components and an interference term. The objective in \cite{boson} is to systematize the procedure like an algorithm and, consequently, to define the soldered action.  The physics considerations will be taken based on the resulting action.  For example, in \cite{Abreu:1998tc}, one of us have obtained a mass generation in the process. 

An iterative Noether procedure was adopted to lift the global
symmetries to local ones. Therefore, we will assume that the symmetries in question are being
described by the local actions $S_{\pm}(\phi_{\pm}^\eta)$, invariant under a
global multi-parametric transformation

\begin{equation}  \label{ii10}
\delta \phi_{\pm}^\eta = \alpha^\eta\;\;,
\end{equation}
where $\eta$ represents the tensorial character of the basic fields in the
dual actions $S_{\pm}$ and, for notational simplicity, will be dropped from
now on. Here the $\pm$ subscript is referring to the opposite/complementary aspects of two models at hand, for instance, $\phi_+$ may refer to a left chiral field and $\phi_-$ to a field with right chirality. As it is well known, we can write,

\begin{equation}
\delta S_{\pm}\,=\,J^{\pm}\,\partial_{\pm}\,\alpha\;\;,
\end{equation}
where $J^{\pm}$ are the Noether currents.

Now, under local transformations these actions will not remain invariant,
and Noether counterterms become necessary to reestablish the invariance,
along with appropriate auxiliary fields $B^{(N)}$, the so-called soldering
fields which have no dynamics where the $N$ superscript is referring to the level of the iteration. This makes a wider range of gauge-fixing conditions available. In this way, the $N$-action can be written as,

\begin{equation}  \label{ii20}
S_{\pm}(\phi_{\pm})^{(0)}\rightarrow S_{\pm}(\phi_{\pm})^{(N)}=
S_{\pm}(\phi_{\pm})^{(N-1)}- B^{(N)} J_{\pm}^{(N)}\;\;.
\end{equation}
Here $J_{\pm}^{(N)}$ are the $N-$iteration Noether currents. For the self
and anti-self dual systems we have in mind that this iterative gauging
procedure is (intentionally) constructed not to produce invariant actions
for any finite number of steps. However, if after N repetitions, the non-invariant piece ends up being only dependent on the gauging parameters, but
not on the original fields, there will exist the possibility of mutual
cancellation if both gauged version of self and anti-self dual systems are put together.
Then, suppose that after N repetitions we arrive at the following
simultaneous conditions,

\begin{eqnarray}  \label{ii30}
\delta S_{\pm}(\phi_{\pm})^{(N)} \neq 0  \nonumber \\
\delta S_{B}(\phi_{\pm})=0\;\;,
\end{eqnarray}
with $S_B$ being the so-called soldered action 
\begin{equation}  \label{ii40}
S_{B}(\phi_{\pm})=S_{+}^{(N)}(\phi_{+}) + S_{-}^{(N)}(\phi_{-})+ %
\mbox{Contact Terms},
\end{equation}
and the "Contact Terms" $\:$ being generally quadratic functions of the soldering
fields. 
Then we can immediately identify the (soldering) interference term as, 
\begin{equation}  \label{ii50}
S_{int}=\mbox{Contact Terms}-\sum_{N}B^{(N)} J_{\pm}^{(N)}\;\;.
\end{equation}
Incidentally, these auxiliary fields $B^{(N)}$ may be eliminated, for instance,
through theirs equations of motion, from the resulting effective action, in
favor of the physically relevant degrees of freedom. It is important to
notice that after the elimination of the soldering fields, the resulting
effective action will not depend on either self or anti-self dual fields $%
\phi_{\pm}$ but only in some collective field, say $\Phi$, defined in terms
of the original ones in a (Noether) invariant way

\begin{equation}
S_{B}(\phi _{\pm })\rightarrow S_{eff}(\Phi )\;\;.  \label{ii60}
\end{equation}
Analyzing in terms of the classical degrees of freedom, it is obvious that we have now a theory with bigger symmetry groups. Once such effective action has been established, the physical consequences of the soldering are readily obtained by simple inspection.

\section{Soldering of NC bosonized Chiral Schwinger model}

The Chiral Schwinger model (CSM) is a 2D (1 spatial dimension + 1 time dimension) Euclidean quantum electrodynamics for a Dirac fermion. This model exhibits a spontaneous symmetry breaking of the U(1) group due to a chiral condensate from a pool of instantons \cite{schwinger}. The photon in this model becomes a massive particle at low temperatures. This model can be solved exactly and it is used as a toy model for other complex theories. The bosonization of this theory can be done in several ways that apparently leads to different bosonized models. But these apparently inequivalent models are related by some gauge transformations \cite{Harada}. Here we shall not enter into the details of this equivalence.  We will discuss the application of the soldering mechanism in the different forms concerning these chiral models.

The CSM is described by the Lagrangian density 
\be \label{ch}
\mathcal{L}_{ch}=\dot{\phi}\phi^{\prime}-(\phi^\prime)^2+2e\phi^\prime(A_0-A_1)-\frac{1}{2}e^2(A_0-A_1)^2+\frac{1}{2}e^2aA_\mu A^\mu \,\,,
\ee

\ni where the last term is the CSM mass term for the gauge field $A_{\mu}$.

In fact, this Lagrangian is the gauged version of the FJ's Lagrangian, $\mathcal{L}_0=\dot{\phi}\phi^{\prime}-(\phi^\prime)^2$ \cite{jackiw}.\\
On the 2D extended Minkowski spacetime $(\tau, x)$ the Lagrangian (\ref{ch}) takes the following action form
\bn
\hat{\mathcal{S}}&=&\int d\tau dx\Bigg[\frac{\p \phi}{\p \tau}\frac{\p \phi}{\p x}- (\frac{\p \phi}{\p x})^2+2e\frac{\p \phi}{\p x}(A_0-A_1) -\frac{1}{2}e^2(A_0-A_1)^2 \\ \no &+&\frac{1}{2}e^2a\eta^{\mu\nu}A_\mu A_\nu-\frac{1}{4}F_{\mu\nu}F^{\mu\nu}\Bigg] \,\,, 
\en

\ni where $\eta^{\mu\nu}=diag(1,-1)$ is the flat metric of the extended Minkowski spacetime  $(\tau,x)$ and $a$ is a real parameter $(a>1)$.

By the coordinate transformation (\ref{trans}) we can rewrite the above action in terms of $(t,x)$ with explicit noncommutativity,
\bn
\hat{\mathcal{S}}&=&\int dt dx \sqrt{-g} \Bigg[\frac{1}{\dot{\tau}}\frac{\p \phi}{\p t}\frac{\p \phi}{\p x} - (\frac{\p \phi}{\p x})^2+2e\frac{\p \phi}{\p x}(A_0-A_1)-\frac{1}{2}e^2(A_0-A_1)^2 \no \\  &+& \frac{1}{2}e^2a\eta^{\mu\nu}A_\mu A_\nu+\frac{1}{2}\left(\frac{1}{\dot{\tau}}\frac{\p A_1}{\p t}-\frac{\p A_0}{\p x} \right) ^2  \Bigg],
\en
where $\sqrt{-g}$ is the Jacobian of the transformation and also the non-trivial measure of the $k$-deformed Minkowski spacetime. Note that always $\sqrt{-g}=\mid \dot{\tau} \mid$ but here we only focus on the case $\dot{\tau}>0$.

Until here we have considered only left chiral Schwinger model but the bosonization process gives us both the left and right chiral bosons which depends on the ''chiral constraint"$\:$ that we have imposed on it. The corresponding Lagrangians for these chiral models in the extended Minkowski spacetime are given by  
\begin{eqnarray}
\hat{\mathcal{L}}_+&=&\dot{\phi}\phi^{\prime}-\sqrt{-g}(\phi^{\prime})^2+\sqrt{-g}\lbrace2e\phi^{\prime}(A_0-A_1)-\frac{1}{2}e^2(A_0-A_1)^2+\frac{1}{2}e^2a[(A_0)^2-(A_1)^2]\rbrace \nonumber\\
&+&\frac{1}{2\sqrt{-g}}(\dot{A_1}-\sqrt{-g}A_0^{\prime})^2 \label{lcsm+}
\end{eqnarray}
\begin{eqnarray}
\hat{\mathcal{L}}_-&=&-\dot{\rho}\rho^{\prime}-\sqrt{-g}(\rho^{\prime})^2+\sqrt{-g}\lbrace2e\rho^{\prime}(A_0-A_1)-\frac{1}{2}e^2(A_0-A_1)^2 +\frac{1}{2}e^2b[(A_0)^2-(A_1)^2]\rbrace  \nonumber\\ &+&\frac{1}{2\sqrt{-g}}(\dot{A_1}-\sqrt{-g}A_0^{\prime})^2 \label{lcsm-}.
\end{eqnarray}
Note that $+$ and $-$ signs are associated to left and right moving chiral bosons, respectively. These models contain noncommutativity through the proper time $\tau$ with the finite NC parameter $k$. In the limit $k\rightarrow +\infty, \sqrt{-g}=\dot{\tau}=1$ these Lagrangians turn back to theirs ordinary forms on the Minkowski spacetime. 

Now we are ready to sold these two chiral Lagrangians. To accomplish the task we calculate the variations of Eqs. (\ref{lcsm+}) and (\ref{lcsm-}) under the following local variations
\begin{equation}
\delta\phi=\eta(t,x)=\delta\rho.
\end{equation}
In fact we are imposing this local symmetry into these models in order to obtain a gauge invariant Lagrangian. Under this variation we have, after some algebra, that
\begin{equation}
\delta(\hat{\mathcal{L}}_+ +\hat{\mathcal{L}}_-)=(J_+ +J_-)\delta B_1
\end{equation}
where
\begin{equation}
J_+=2\dot{\phi}-2\sqrt{-g}\phi^{\prime}+2e\sqrt{-g}(A_0-A_1)
\end{equation}
and
\begin{equation}
J_-=-2\dot{\rho}-2\sqrt{-g}\rho^{\prime}+2e\sqrt{-g}(A_0-A_1)
\end{equation}

\ni where $B_1$ (mentioned in the previous section) and $B_2$ (which will be necessary) are auxiliary fields which variations can be defined as
\begin{equation}
\delta B_1=\partial_x\eta \hspace{0.4cm} \text{and} \hspace{0.4cm} \delta B_2=\partial_t \eta.
\end{equation} 

So we must add a counterterm to both original Lagrangians (\ref{lcsm+}) and (\ref{lcsm-}) to cover the above extra terms. So
\begin{eqnarray}
\hat{\mathcal{L}}_{+1}&=&\hat{\mathcal{L}}_{+}-J_+B_1 \hfill \\
\hat{\mathcal{L}}_{-1}&=&\hat{\mathcal{L}}_{-}-J_-B_1 
\end{eqnarray}
Now let us check the variation of the above Lagrangians 
\begin{eqnarray}
\delta \hat{\mathcal{L}}_{+1}&=&-(\delta J_+)B_1=-(2\dot{\eta}-2\sqrt{-g}\eta^{\prime})B_1 \nonumber \hfill \\
&=&-2B_1(\delta B_2)+2\sqrt{-g}B_1(\delta B_1)  \hfill \\
\delta \hat{\mathcal{L}}_{-1}&=&-(\delta J_-)B_1=(2\dot{\eta}-2\sqrt{-g}\eta^{\prime})B_1 \nonumber \hfill \\ &=&2B_1(\delta B_2)+2\sqrt{-g}B_1(\delta B_1).
\end{eqnarray}
As we can see, it is not zero but the extra terms are independent of original fields. So the iteration will finish in this second step by adding another counterterm.

Finally we can sold these two Lagrangians in order to construct an invariant one. 
\begin{equation}
\hat{\mathcal{W}}=\hat{\mathcal{L}}_++\hat{\mathcal{L}}_--(J_++J_-)B_1-2\sqrt{-g}(B_1)^2.
\end{equation} 

\ni where the $B_2$ field were eliminated algebraically.  On the other hand,  we can eliminate the auxiliary field $B_1$ by its equation of motion
\begin{equation}
\frac{\delta \mathcal{W}}{\delta B_1}=0\Longrightarrow-(J_++J_-)-4\sqrt{-g}B_1=0\Rightarrow B_1=\frac{-1}{4\sqrt{-g}}(J_++J_-) \label{bcsm}
\end{equation}
By substituting Eq. (\ref{bcsm}) into $\mathcal{W}$ we find
\begin{equation}
\hat{\mathcal{W}}=\hat{\mathcal{L}}_++\hat{\mathcal{L}}_-+\frac{1}{8\sqrt{-g}}(J_++J_-)^2.
\end{equation}
Here we define a new field the soldering field $\Psi=\phi-\rho$. By this definition we can rewrite $\mathcal{W}$ in a compact and nice form
\begin{equation} \label{csm1}
\hat{\mathcal{W}}=-\frac{\sqrt{-g}}{2}\Psi^{\prime 2}+\frac{1}{2\sqrt{-g}}\dot{\Psi}^2+2e\dot{\Psi}(A_0-A_1)+2 \xi
\end{equation}
where $\xi$ is 
\begin{equation}
\xi=\sqrt{-g}\lbrace\frac{1}{2}e^2(A_0-A_1)^2+\frac{1}{4}e^2(a+b)[(A_0)^2-(A_1)^2]\rbrace+\frac{1}{2\sqrt{-g}}(\dot{A_1}-\sqrt{-g}A_0^{\prime})^2
\end{equation}

As the final result, the action (\ref{csm1}) is not ``chiral"$\:$ theory anymore and it has a bigger symmetry group than the two initial models. To this aim, we have soldered the two chiral models and as a consequence we have gained an additional term in the final Lagrangian that was absent initially. One of the peculiar consequences  of this action is that the electromagnetic field interacts just with the temporal derivative of the soldered field. This peculiarity has its origin in the noncovariant initial Jackiw-Floreanini Lagrangian. In fact one can decompose the above action into two distinct ones using the dual projection approach \cite{}. The result is a self-dual and a free massive scalar fields.      

This mechanism in some sense is analogous to adding a mass term into the Dirac action. Without this mass term the Dirac equation describes two chiral electrons and by adding the mass, we have merged these two chiral electrons to obtain the real electron.  

\section{The soldering of the generalized bosonized CSM}

Bassetto et al. \cite{Bassetto:1993dh} have suggested the generalized chiral Schwinger model
(GCSM), i.e., a vector and axial-vector theory characterized by a parameter which interpolates between pure vector and chiral Schwinger models. This 2D model is given by the action
\begin{equation}
\hat{\mathcal{S}} = \int \mathrm{d}t \mathrm{d}x\left[ \frac{1}{2}\left( {\partial {}_\mu \phi } \right)\left( {{\partial ^\mu }\phi } \right)
+ e{A_\mu }\left( {{\epsilon ^{\mu \nu }} - r{{\eta}^{\mu \nu }}} \right){\partial_\nu }\phi
+ \frac{1}{2}{e^2}a{A_\mu }{A^\mu } - \frac{1}{4}{F_{\mu \nu }}{F^{\mu \nu }} \right].
\end{equation}
The quantity $r$ is a real interpolating parameter between the vector $(r = 0)$ and the chiral Schwinger models $(r = \pm 1)$. This action can be rewritten in the extended Minkowski spacetime 
\begin{eqnarray}
\hat{\mathcal{L}}=\frac{1}{2\sqrt{-g}}\dot{\phi}^2-\frac{\sqrt{-g}}{2}\phi^{\prime 2}-k_1\dot{\phi}+k_2\phi^{\prime}+\xi
\end{eqnarray}
where
\begin{eqnarray}
k_1&=&e(rA_0+A_1) \hfill \\
k_2&=&e\sqrt{-g}(A_0+rA_1) \hfill \\
\xi &=&\frac{1}{2\sqrt{-g}}\left(\dot{A}_1-\sqrt{-g}A^{\prime}_0\right)^2+\frac{1}{2}e^2a\sqrt{-g}\left[ (A_0)^2-(A_1)^2\right]. 
\end{eqnarray}
By defining the value of the parameter $r$ in two extreme points $\pm1$ we obtain two chiral Lagrangians
\begin{eqnarray}
\hat{\mathcal{L}}_+&=&\frac{1}{2\sqrt{-g}}\dot{\phi}^2-\frac{\sqrt{-g}}{2}\phi^{\prime 2} -e(A_0+A_1)\dot{\phi}+e\sqrt{-g}(A_0+A_1)\phi^{\prime} \hfill \nonumber \\
&+&\frac{1}{2\sqrt{-g}}\left(\dot{A}_1-\sqrt{-g}A_0^{\prime} \right)^2 +\frac{1}{2}ae^2\sqrt{-g}\left[ (A_0)^2-(A_1)^2\right]  \hfill \label{gcsm1}\\
\hat{\mathcal{L}}_-&=&\frac{1}{2\sqrt{-g}}\dot{\rho}^2-\frac{\sqrt{-g}}{2}\rho^{\prime 2} -e(-A_0+A_1)\dot{\rho}+e\sqrt{-g}(A_0-A_1)\rho^{\prime} \hfill \nonumber \\
&+&\frac{1}{2\sqrt{-g}}\left(\dot{A}_1-\sqrt{-g}A_0^{\prime} \right)^2 +\frac{1}{2}be^2\sqrt{-g}\left[ (A_0)^2-(A_1)^2\right] \label{gcsm2}
\end{eqnarray}

\ni where $a$ and $b$ are the Jackiw-Rajaraman coefficients for each chirality, respectively. Here, through the  iterative Noether embedding procedure, we will transform both Lagrangians (\ref{gcsm1}) and (\ref{gcsm2}) into two embedded Lagrangians which are invariant under transformations $\delta\phi=\eta(x)$ and $\delta\rho=\eta(x)$. After that, we will be able to sold these new Lagrangians in order to yield an invariant one that describes a fermionic system. By varying the Lagrangians with respect to the variables $\p_t \Phi$ and $\p_x \Phi$, $(\Phi=(\phi, \rho))$, we obtain the following Noether currents
\begin{eqnarray}
J_{1+}&=&\frac{1}{\sqrt{-g}}\dot{\phi}-e(A_0+A_1) \hfill \\
J_{2+}&=&-\sqrt{-g}\left[{\phi}^{\prime}-e(A_0+A_1) \right] \hfill \\
J_{1-}&=& \frac{1}{\sqrt{-g}}\dot{\rho}+e(A_0-A_1)  \hfill \\
J_{2-}&=&-\sqrt{-g}\left[{\rho}^{\prime}-e(A_0-A_1)\right]. 
\end{eqnarray}
After two iterations and by adding the counterterms to the original Lagrangians, we can find that
\begin{eqnarray}
\hat{\mathcal{L}}_{+}^{(2)}&=\mathcal{L}_+-J_{1+}B_1-J_{2+}B_2+\frac{1}{2\sqrt{-g}}(B_1)^2-\frac{\sqrt{-g}}{2}(B_2)^2+\xi_+  \label{lgcsm+}\hfill \\
\hat{\mathcal{L}}_{-}^{(2)}&=\mathcal{L}_--J_{1-}B_1-J_{2-}B_2+\frac{1}{2\sqrt{-g}}(B_1)^2-\frac{\sqrt{-g}}{2}(B_2)^2 +\xi_-  \label{lgcsm-}
\end{eqnarray} 
where $\xi_\pm$ are non-dynamical terms of $\mathcal{L}_{\pm}$. The embedding process ends after these two steps and these Lagrangians are invariant under the desired transformation $\delta\phi=\eta(x)\delta\rho$. 
Now we can solder them by adding up two Lagrangians Eqs. (\ref{lgcsm+}) and (\ref{lgcsm-})
\begin{eqnarray}
\hat{\mathcal{W}}&=&\hat{\mathcal{L}}_{+}^{(2)}+\hat{\mathcal{L}}_{-}^{(2)} \hfill \\
&=&\hat{\mathcal{L}}_++\hat{\mathcal{L}}_--(J_{1+}+J_{1-})B_1-(J_{2+}+J_{2-})B_2+\frac{1}{\sqrt{-g}}(B_1)^2-\sqrt{-g}(B_2)^2 \nonumber
\end{eqnarray}
To express this Lagrangian just in terms of the original fields, we can eliminate $B_1$ and $B_2$ easily by using their equation of motions, which reads
\begin{eqnarray}
B_1&=\frac{\sqrt{-g}}{2}(J_{1+}+J_{1-}) \hfill \\ 
B_2&=-\frac{1}{2\sqrt{-g}}(J_{2+}+J_{2-}).
\end{eqnarray}
After substituting these results into $\mathcal{W}$, defining a new field $\Psi=\phi-\rho$ and fixing the Jackiw-Rajaraman coefficients $a=b=1$, for simplicity, we can write that
\begin{eqnarray}
\hat{\mathcal{W}}&=&\frac{1}{4\sqrt{-g}}\dot{\Psi}^2- \frac{\sqrt{-g}}{4}\Psi^{\prime2}-eA_0\dot{\Psi}+eA_1\sqrt{-g}\Psi^{\prime} \hfill \nonumber \\
&+&\frac{1}{2\sqrt{-g}}\left(\dot{A}_1-\sqrt{-g}A^{\prime}_0\right)^2+e^2\sqrt{-g}\left[ (A_0)^2-(A_1)^2\right]. 
\end{eqnarray}
This Lagrangian describes a 2D fermionic system and has a larger symmetry group than the initial Lagrangians (\ref{gcsm1}) and (\ref{gcsm2}). As the previous case, the soldering process included an extra noton term into the original Lagrangians to fuse the chiral states. This non-dynamical term can acquire dynamics upon quantization \cite{boson}.

\section{The soldering of the gauge invariant generalized bosonized CSM}

In \cite{Miao:1996xs}, the authors have introduced the Wess-Zumino (WZ) term for the GCSM and
constructed its gauge invariant formulation by adding the WZ term into the Lagrangian of the model. This gauge invariant model is described by
\begin{eqnarray}
\hat{\mathcal{S}} &=&\int{\mathrm{d}t \mathrm{d}x}\left\{\frac{1}{2}\left( {\partial {}_\mu \phi } \right)\left( {{\partial ^\mu }\phi } \right)
+ e{A^\mu }\left( {{\epsilon _{\mu \nu }} - r{\eta_{\mu \nu }}} \right){\partial ^\nu }\phi +\frac{1}{2}{e^2}a{A_\mu }{A^\mu }
- \frac{1}{4}{F_{\mu \nu }}{F^{\mu \nu }} \right.\nonumber \\
& &\left.+ \frac{1}{2}\left( {a - {r^2}} \right)\left( {\partial {}_\mu \theta }\right)\left( {{\partial ^\mu }\theta } \right)
+e{A^\mu }\left[ {r{\epsilon _{\mu \nu }} + \left( {a - {r^2}} \right){\eta_{\mu \nu }}} \right]{\partial ^\nu }\theta \right\},
\end{eqnarray}

\ni where $\theta(x)$ is the WZ field. The Lagrangians of left/right moving bosons are given by defining the parameter $r$ at its two opposite points $\pm 1$
\begin{eqnarray}
\hat{\mathcal{L}}_+&=&\frac{1}{2\sqrt{-g}}(\dot{\phi})^2-\frac{\sqrt{-g}}{2}(\phi^{\prime})^2-b_1\dot{\phi}+b_1\sqrt{-g}\phi^{\prime} +\frac{b_2}{\sqrt{-g}}(\dot{\theta})^2-b_2\sqrt{-g}(\theta^{\prime})^2+b_3\dot{\theta}+b_4\theta^{\prime}+\xi_+ \nonumber  \\
\hat{\mathcal{L}}_-&=&\frac{1}{2\sqrt{-g}}(\dot{\rho})^2-\frac{\sqrt{-g}}{2}(\rho^{\prime})^2-b_5\dot{\rho}+b_5\sqrt{-g}\rho^{\prime}
+\frac{b_2^\prime}{\sqrt{-g}}(\dot{\eta})^2-b_2^\prime\sqrt{-g}(\eta^{\prime})^2+b_6\dot{\eta}+b_7\eta^{\prime}+\xi_- 
\end{eqnarray}
where $\eta$ is also another WZ field and
\begin{eqnarray}
b_1 &\equiv& e(A_0+A_1), \hspace{.7cm} b_2\equiv\frac{a-1}{2}, \hspace{.7cm} b_2^\prime\equiv\frac{b-1}{2}, \hspace{.7cm} b_3\equiv e\left[ A_0(a-1)-A_1\right] \no \\ 
b_4 &\equiv& e\sqrt{-g}\left[A_0-A_1(a-1) \right], \hspace{.7cm}
b_5 \equiv e(A_0-A_1), \hspace{.7cm} \nonumber  \\ 
b_6 &\equiv& e\left[ A_0(b-1)+A_1\right], \hspace{.7cm} b_7\equiv e\sqrt{-g}\left[-A_0-A_1(b-1) \right] \nonumber  \\
\xi_\pm &\equiv& \frac{1}{2\sqrt{-g}}(\dot{A_1})^2-\frac{\sqrt{-g}}{2}(A_0^{\prime})^2+\frac{\sqrt{-g}}{2}e^2\left(^a_b \right)\left[(A_0)^2-(A_1)^2 \right]-\dot{A_1}A_0^{\prime}. 
\end{eqnarray} 

The goal here is to gauge these Lagrangians under the following transformations
\begin{eqnarray}\label{trans-gigcsm}
\delta\phi=\delta\rho=\alpha(x) \no\hfill \\
\delta\theta=\delta\eta=\beta(x). 
\end{eqnarray}
The Noether currents under these transformations are
\begin{eqnarray}
J_{1+} &=&\frac{1}{\sqrt{-g}}\dot{\phi}-b_1, \hspace{.7cm}\hspace{1.2cm}
J_{1-} =\frac{1}{\sqrt{-g}}\dot{\rho}-b_5, \hspace{.7cm} \no \\ 
J_{2+} &=&-\sqrt{-g}\phi^{\prime}+b_1\sqrt{-g}, \hspace{.7cm}  
J_{2-} =-\sqrt{-g}\rho^{\prime}+b_5\sqrt{-g}, \hspace{.7cm} \no \\ 
J_{3+} &=&\frac{2b_2}{\sqrt{-g}}\dot{\theta}+b_3, \hspace{.7cm} \hspace{1.2cm}
J_{3-} =\frac{2b_2^\prime}{\sqrt{-g}}\dot{\eta}+b_6, \hspace{.7cm} \no \\
J_{4+} &=&-2b_2\sqrt{-g}\theta^{\prime}+b_4,  \hspace{1.2cm}
J_{4-} =-2b_2^\prime\sqrt{-g}\eta^{\prime}+b_7. 
\end{eqnarray}
The first iteration Lagrangians read
\begin{eqnarray} \label{gigscm1}
\hat{\mathcal{L}}_+^{(1)}=\hat{\mathcal{L}}_+-J_{1+}B_1-J_{2+}B_2-J_{3+}B_3-J_{4+}B_4 \nonumber \\
\hat{\mathcal{L}}_-^{(1)}=\hat{\mathcal{L}}_--J_{1-}B_1-J_{2-}B_2-J_{3-}B_3-J_{4-}B_4
\end{eqnarray}
where $B_1, B_2, B_3$ and $B_4$ are new auxiliaries fields which have the following variations
\begin{equation}
\delta B_1=\partial_t \alpha, \hspace{.7cm} \delta B_2=\partial_x \alpha, \hspace{.7cm} \delta B_3=\partial_t \beta, \hspace{.7cm} \delta B_4=\partial_x \beta.
\end{equation} 
The variation of the first iterated Lagrangians are given by
\begin{eqnarray}
\delta \hat{\mathcal{L}}_+^{(1)}=-\frac{1}{\sqrt{-g}}(\delta B_1)B_1+\sqrt{-g}(\delta B_2)B_2-\frac{2b_2}{\sqrt{-g}}(\delta B_3)B_3+2b_2\sqrt{-g}(\delta B_4)B_4 \hfill \\
\delta \hat{\mathcal{L}}_-^{(1)}=-\frac{1}{\sqrt{-g}}(\delta B_1)B_1+\sqrt{-g}(\delta B_2)B_2-\frac{2b_2}{\sqrt{-g}}(\delta B_3)B_3+2b_2\sqrt{-g}(\delta B_4)B_4.
\end{eqnarray}

As we can see, these variations are completely independent of the original fields.  Therefore the embedding process finished here and by adding the counterterms associated with these variations we can obtain our desired invariant Lagrangian. Now we are ready to fuse both Lagrangians in Eqs. (\ref{gigscm1}) by adding them up and introducing a counterterm
\begin{eqnarray}
\hat{\mathcal{W}}&=&\hat{\mathcal{L}}_++\hat{\mathcal{L}}_--J_{1+}B_1-J_{2+}B_2-J_{3+}B_3-J_{4+}B_4-J_{1-}B_1-J_{2-}B_2-J_{3-}B_3-J_{4-}B_4 \nonumber \hfill\\
&+&\frac{1}{\sqrt{-g}}(B_1)^2-\sqrt{-g}(B_2)^2+\frac{2b_2}{\sqrt{-g}}( B_3)^2-2b_2\sqrt{-g}(B_4)^2 \,\,,\label{winvar}
\end{eqnarray}
where we have fixed the Jackiw-Rajaraman coefficients $a=b$ for simplicity. To express the final result only in terms of the original fields, one can eliminate the auxiliary fields by using their equations of motions
\begin{eqnarray}
B_1&=&\frac{\sqrt{-g}}{2}(J_{1+}+J_{1-}) \hfill \\
B_2&=&\frac{-1}{2\sqrt{-g}}(J_{2+}+J_{2-}) \nonumber \hfill \\
B_3&=&\frac{\sqrt{-g}}{4b_2}(J_{3+}+J_{3-}) \nonumber \hfill \\
B_4&=&\frac{-1}{4b_2\sqrt{-g}}(J_{4+}+J_{4-}). \nonumber
\end{eqnarray}
By substituting these results into Eq. (\ref{winvar}) and introducing two soldering fields $\Psi=\phi-\rho$ and $\Omega=\theta-\eta$ we obtain an effective action
\begin{eqnarray}
\hat{\mathcal{W}}_{eff}&=&\frac{1}{4\sqrt{-g}}(\dot{\Psi})^2-\frac{\sqrt{-g}}{4}(\Psi^{\prime})^2-eA_0\dot{\Psi}+e\sqrt{-g}A_1\Psi^{\prime}+\frac{b_2}{2\sqrt{-g}}(\dot{\Omega})^2 \hfill \\
&-&\frac{b_2\sqrt{-g}}{2}(\Omega^{\prime})^2-eA_1\dot{\Omega} +\frac{1}{2}\left[eA_0+e\sqrt{-g}(A_0-2A_1b_2)+2eA_1b_2 \right]\Omega^{\prime} \nonumber \hfill \\
&-&2e^2b_2\sqrt{-g}(A_0)^2+\frac{e^2\sqrt{-g}}{8b_2}\left(A_0-2A_1b_2 \right)^2-\frac{2e^2A_0}{8b_2}(A_0-2A_1b_2)\nonumber \hfill \\
&+&\frac{e^2}{8\sqrt{-g}b_2}(A_0)^2+\frac{e^2}{2\sqrt{-g}}A_0A_1+\frac{e^2b_2}{2\sqrt{-g}}(A_1)^2-\frac{e^2}{2}A_1(A_0-2A_1b_2) +2\xi \no
\end{eqnarray}
where $\xi=\xi_-+\xi_+$. The initial Lagrangians were invariant under a semilocal gauge group, but this effective Lagrangian is invariant under the local version of the initial gauge group and moreover it is invariant under gauge transformations (\ref{trans-gigcsm}).

One can ask about the counterpart of this model in the commutative spacetime. We can find it just by putting $\sqrt{-g}=1$. It reads
\bn
\mathcal{W}_{eff}&=&\frac{1}{4}\p_\mu \Psi \p^\mu \Psi+e\ep^{\mu\nu}A_\mu \p_\nu\Psi +\(a-1\)\p_\mu \Omega \p^\mu \Omega-eA_\mu \epsilon^{\mu\nu}\p_\nu \Omega \\ 
&+& \frac{1}{2}{e^2}a{A_\mu }{A^\mu } - \frac{1}{4}{F_{\mu \nu }}{F^{\mu \nu }} \no  
\en
where $\xi^\prime=\xi|_{\sqrt{-g}=1}$. We have succeeded in including the effects of interference between rightons and leftons (right/left moving scalar). Consequently, these components have lost their individuality in favor of a new, gauge invariant, collective field that does not depend on $\phi$ or $\rho$ separately.

As it can be seen, this Lagrangian is apparently different from the initial ones and the new fields $\Psi$ and $\Omega$ are not chiral anymore. If we fix the Jackiw-Rajaraman coefficients $a=b=1$, the field $\Omega$ becomes non-dynamical and it will just interact with electromagnetic field. The combination of the massless modes led us to a massive vectorial mode as a consequence of the chiral interference. The noton field, that was defined before, propagates neither to the left nor to the right directions.

\section{The soldering of NC (anti)self-dual models in D=2+1}

The Thirring model is an exactly solvable QFT that describes the self-interactions of a Dirac theory in (2+1) dimensions. For the first time S. Coleman has discovered an equivalence between this model and the Sine-Gordon one which is a bosonic theory \cite{coleman}.

In $D=1+1$, the starting point was to consider two distinct fermionic theories with opposite chiralities. The analogous thing is to take two independent Thirring models with identical coupling strengths but opposite mass signatures,
\bn
\label{240}
{\mathcal{L}_+}&=&\bar\psi\left(i\dslash + m\right)\psi -\frac{\lambda^2}{2}
\left(\bar\psi\gamma_\mu\psi\right)^2\nonumber\\
{\mathcal{L}_-} &=& \bar \xi\left(i\dslash - m'\right)\xi - \frac{\lambda^2}{2} 
\left(\bar\xi\gamma_\mu\xi\right)^2 \,\,,
\en

\ni  where the bosonized Lagrangians are, respectively,
\bn
\label{250}
{\mathcal{L}_+}&=&\frac{1}{2M} 
\epsilon_{\mu\nu\lambda}f^\mu\partial^\nu f^\lambda +
{1\over 2} f_\mu f^\mu\nonumber\\
{\mathcal{L}_-} &=&- \frac{1}{2M}
\epsilon_{\mu\nu\lambda}g^\mu\partial^\nu g^\lambda +
{1\over 2} g_\mu g^\mu, 
\en
where $f_\mu$ and $g_\mu$ are the distinct bosonic vector fields.  The current
bosonization formula in both cases are given by

\bn
\label{255}
j_\mu^+ &=& \bar\psi\gamma_\mu\psi=\frac{\lambda}{4\pi} 
\epsilon_{\mu\nu\rho}\partial^\nu f^\rho\nonumber\\
j_\mu^- &=&\bar\xi\gamma_\mu\xi= - \frac{\lambda}{4\pi} 
\epsilon_{\mu\nu\rho}\partial^\nu g^\rho.
\en
These models are known as the self and anti-self dual models \cite{Townsend:1983xs, Banerjee:1999mc, Banerjee:1995yf}. 

On the extended Minkowski spacetime $(\tau, x)$ the Lagrangian (\ref{250}) takes the following action form
\be
\hat{S}_\pm=\int d\tau d^2x \left[ \frac{1}{2}h^\mu h_\mu \pm\frac{1}{2M}\left( \ep_{\mu 0 \lambda}h^\mu \frac{\p h^{\lambda}}{\p\tau} + \ep_{\mu i \lambda} h^\mu\p^ih^\lambda\right) \right] 
\ee
where $h^\mu=f^\mu, g^\mu$.

After making the coordinate transformation, we obtain the action written in terms of the coordinates $(t, x)$,
\be
\hat{S}_\pm=\int dt d^2x \sqrt{-g} \left[\frac{1}{2}h^\mu h_\mu \pm \frac{1}{2M} \ep_{\mu i \lambda}h^\mu \p^i h^\lambda \right] \pm \frac{1}{2M} \ep_{\mu 0 \lambda} h^\mu \frac{\p h^\lambda}{\p t}.
\ee
Taking a hint from the two dimensional case, let us consider the gauging of the following symmetry
\be \label{var3d}
\de f_\mu=\de g_\mu= \ep_{\mu \rho \sigma}\p^\rho \al^\sigma.
\ee 
Under these transformations the bosonized Lagrangians change as

\begin{eqnarray} \label{3dvar-first}
\delta \hat{S}_\pm &=& \int dt d^2x \left[ \sqrt{-g} \left\lbrace  \epsilon^{\mu \rho \sigma} h_\mu \pm \frac{1}{M} \epsilon_{\mu i\lambda} \epsilon^{\mu \rho \sigma} \partial^{i} h^{\lambda} \right\rbrace \pm \frac{1}{M} \ep_{\mu 0 \lambda} \ep^{\mu \rho \sigma} \p^0 h^{\lambda}\right] \p_\rho \al_\sigma.
\end{eqnarray}

\ni We can identify the Noether currents 

\bn \label{curent3d}
J_{\pm}^{\rho \sigma}(h_{\mu})=\sqrt{-g} \left\lbrace  \epsilon^{\mu \rho \sigma} h_\mu \pm \frac{1}{M} \epsilon_{\mu i\lambda} \epsilon^{\mu \rho \sigma} \partial^{i} h^{\lambda} \right\rbrace \pm \frac{1}{M} \ep_{\mu 0 \lambda} \ep^{\mu \rho \sigma} \p^0 h^{\lambda}.
\en 

As a comment about the form of the gauge transformation in Eq. (\ref{var3d}) we can say that the simpler form such as the one we have assumed in 2D case, will not be suitable and the variations cannot be combined to give a single structure like the (\ref{curent3d}) one. Now we introduce the auxiliary field coupled to the antisymmetric currents. In the two dimensional case, this field was a vector. In the three dimensional case, as a natural generalization, we adopt an antisymmetric second rank Kalb-Ramond tensor field $B_{\rho\sigma}$ where its transformation is given by
\be \label{var3db}
\de B_{\rho\sigma}= \p_\rho \al_\sigma - \p_\sigma \al_\rho
\ee

It is worthwhile to mention that in the canonical NC approach, one must include the variation of the current associated with the NC field/parameter concerning the transformation of the auxiliary tensor field in order to obtain an effective Lagrangian after the soldering procedure \cite{Ghosh:2003yt}. 

To eliminate the non-vanishing change (\ref{3dvar-first}), we add a counter-term to the original Lagrangian. So, the first iterated Lagrangians are
\be
\mathcal{L}_{\pm}^{(1)}= \mathcal{L}_{\pm} - \frac{1}{2} J_{\pm}^{\rho \sigma} (h_{\mu}) B_{\rho\sigma}
\ee
which transforms as,
\be
\de \mathcal{L}_\pm^{(1)}= -\frac{1}{2} \de J_{\pm}^{\rho \sigma} B_{\rho\sigma}.
\ee
The variation of the currents coupled to the auxiliary field is
\be
\de J^{\rho\sigma}_{\pm} B_{\rho \sigma} = \sqrt{-g}\Big [\de B^{\rho\sigma}B_{\rho \sigma} \mp \frac{1}{M} \ep^{\lambda \ga \ta} (\p^i\p_\ga \al_\ta) B_{i\lambda} \Big ] \mp \frac{2}{M}\ep^{\lambda \ga \ta} (\p^0 \p_\ga \al_\ta)B_{0\lambda} . 
\ee
As we can see, the above Lagrangians also are not invariant under the transformations (\ref{var3d}), hence we must go further and add another counter term. As a key point in the soldering formalism, the invariance of one Lagrangian alone is not desired. We are looking for a combination of both Lagrangians that are gauge invariant. To this aim, the second iteration Lagrangians is defined by
\be
\mathcal{L}_\pm^{(2)}= \mathcal{L}_\pm^{(1)} + \frac{\sqrt{-g}}{4}B^{\rho\sigma} B_{\rho\sigma}.
\ee  
By this definition, a straightforward algebra shows that the following combination is invariant under transformation (\ref{var3d}) and (\ref{var3db}).  So,
\bn \label{sold1}
\mathcal{L}_S &=& \mathcal{L}_+^{(2)}+\mathcal{L}_-^{(2)} \no \\
&=& \mathcal{L}_+ + \mathcal{L}_- - \frac{1}{2}B^{\rho \sigma} \left(J^+_{\rho \sigma}(f) + J^-_{\rho \sigma}(g) \right) + \frac{\sqrt{-g}}{2} B^{\rho\sigma} B_{\rho\sigma}.
\en
The gauging procedure of the symmetry is therefore complete now. But the final result would be more interesting if we express the above Lagrangian in terms of the original fields. By using the equation of motion for $B_{\rho \sigma}$ we can eliminate this auxiliary field
\be
B_{\rho \sigma}= \frac{1}{2\sqrt{-g}} \left( J^+_{\rho \sigma}(f) + J^-_{\rho \sigma}(g) \right). 
\ee  
Including this solution into (\ref{sold1}) the final soldered Lagrangian is expressed only in terms of the original fields,
\be
\mathcal{L}_S= \mathcal{L}_+ + \mathcal{L}_- - \frac{1}{8\sqrt{-g}} \left( J^+_{\rho \sigma}(f) + J^-_{\rho \sigma}(g) \right) \left( J^{+\rho \sigma}(f) + J^{-\rho \sigma}(g) \right). 
\ee
The crucial point of soldering formalism becomes clear now.   By using the explicit structures for the currents, the above Lagrangian is no longer a function of $f_\mu$ and $g_\mu$ separately, but solely on the combination
\be
A_\mu=\frac{1}{\sqrt{2}M}(f_\mu - g_\mu).
\ee
By this field redefinition we can obtain the final effective action as
\be \label{proca3d}
\mathcal{L}_S= \frac{M^2\sqrt{-g}}{2}A^\mu A_\mu +\p_i A_0 \p^0 A^i - \frac{1}{2\sqrt{-g}}\p_0 A_i \p^0 A^i - \frac{\sqrt{-g}}{2}(\p^i A^0 \p_i A_0 + \p_i A_j \p^i A^j - \p_j A_i \p^i A^j).
\ee
In the usual commutative Minkowski spacetime we yield the Proca theory by soldering two (anti)self-dual theories \cite{Banerjee:1995yf}. As a generalization, we claim that the Lagrangian (\ref{proca3d}) are the NC version of the Abelian Proca theory in the $\kappa$-deformed (2+1)D Minkowski spacetime. In order to check that our calculation is correct we can obtain directly this Lagrangian by applying the coordinate transformation $(\tau, x)\rightarrow (t,x)$ in Proca theory. The Abelian Proca model on the extended Minkowski spacetime $(\tau, x)$ is 
\bn
\hat{S}&=&\int d\tau d^2x \Big[ -\frac{1}{4}F^{\mu\nu}F_{\mu\nu} + \frac{M^2}{2}A^\mu A_\mu \big] \no \\
&=& \int d\tau d^2x \left(-\frac{1}{2} \big[ \frac{\p A^i}{\p \tau}( \frac{\p A_i}{\p \tau}-\frac{\p A_0}{\p x^i}) + \frac{\p A^0}{\p x_i} (\frac{\p A_0}{\p x^i}-\frac{\p A_i}{\p \tau}) 
+ \frac{\p A^j}{\p x_i}(\frac{\p A_j}{\p x^i} - \frac{\p A_i}{\p x^j}) \big] + \frac{M^2}{2} A^\mu A_\mu \right) \nonumber \\
\en 

\ni where $F^{\mu \nu}= \p^\mu A^\nu - \p^\nu A^\mu$. 
By a coordinate transformation (\ref{trans}) we can rewrite the above actions in terms of $(t,x)$ with explicit noncommutativity,
\bn
\hat{\mathcal{S}}&=&\int dt d^2x \sqrt{-g} \Big(-\frac{1}{2} \Big[ \frac{1}{\sqrt{-g}}\frac{\p A^i}{\p t}( \frac{1}{\sqrt{-g}}\frac{\p A_i}{\p t}-\frac{\p A_0}{\p x^i}) + \frac{\p A^0}{\p x_i} (\frac{\p A_0}{\p x^i}-\frac{1}{\sqrt{-g}}\frac{\p A_i}{\p t}) \no \\ 
&+& \frac{\p A^j}{\p x_i}(\frac{\p A_j}{\p x^i} - \frac{\p A_i}{\p x^j}) \Big] + \frac{M^2}{2} A^\mu A_\mu \Big).
\en
Here we have assumed that $\dot{\tau}=\sqrt{-g}>0$. After some straightforward manipulation we find that
\bn
\hat{\mathcal{S}}&=&\frac{1}{2}\int dt d^2x \Big(  2\p^0 A^i \p_i A_0 + \sqrt{-g}\p^iA^j\p_j A_i - \sqrt{-g}\p^iA^j\p_i A_j \no \\ &-& \sqrt{-g}\p^iA^0\p_i A_0 - \frac{1}{\sqrt{-g}} \p^0A^i\p_0 A_i + M^2\sqrt{-g} A^\mu A_\mu\Big).
\en
As we have expected, this action is equal to the model described by the Lagrangian (\ref{proca3d}). 

Notice that this NC version is that, besides the modification of the field dynamics in this new spacetime, the mass term has also changed and it is not equal to the usual Minkowski spacetime so, the particle associated with this field must have a different mass in this spacetime.

It is noteworthy that the transformations (\ref{var3d}) are not the unique ones that lead to this result. We can also use the transformation

\be \label{var3d-2}
\de f_\mu=-\de g_\mu= \ep_{\mu \rho \sigma}\p^\rho \al^\sigma.
\ee 
By assuming the above transformation and defining the final soldered field 
\be
A_\mu=\frac{1}{\sqrt{2}M}(f_\mu - g_\mu)
\ee

\ni we can arrive at the same Lagrangian as in (\ref{proca3d}). This result led the authors of \cite{gs} to the idea of generalizing the soldering formalism. As it was mentioned before, the basic idea of soldering was that adding two independent dual Lagrangians does not give us new information and for obtaining a gauge invariant model we have to fuse two Lagrangians via the Noether procedure. This idea was successfully applied to different models in various dimensions such as chiral Schwinger model with opposite chiralities. 

Some years after proposing this idea it was shown that the usual sum of opposite chiral bosons models is, in fact, gauge invariant and it corresponds to a composite model, where the component models are the vector and axial Schwinger models \cite{gs}. As a consequence, we can reinterpret the soldering formalism as a kind of degree of freedom reduction mechanism.

In the case at hand, two transformations (\ref{var3d}) and (\ref{var3d-2}) result in the same effective action but in a general case we may obtain two apparently different actions. For example, if we add an interaction term to the Lagrangians (\ref{250}), the final result will be different. This property is the subject of the generalized soldering formalism \cite{gs}. Now this question may arise whether these two actions are describing two distinct phenomena. However, by calculating the generating functional of these two Lagrangians we have the same result. This shows that we are dealing with the same physics but described by different Lagrangians.

\section{Conclusions and perspectives}

The idea that we can construct a bosonized version for some fermionic models in order to study the properties of the target model through a theoretically easier version, the bosonized one, has dwelled in the theoretical physicists mind during the 80's and 90's.  In two spacetime dimensions, the concept of chirality together with the bosonization one were discussed after the influence of the chiral boson version in string theories.

Holding that thought, M. Stone provided a method which objective was to put together in the same multiplet, two chiral versions of bosonized model in such a way that an effective final model was obtained and the target was to analyze physically the properties of this last one.  Another result obtained in the soldering technique is to discuss the fact that the final action is connected to the first ones through duality properties.

There was a relevant production of papers considering several models but none of them have considered NC models, which in fact was our objective here.  We have analyzed the $\kappa$-Minkowski noncommutativity where some variations of the CSM were soldered and the soldered (final) action yielded were discussed in the aftermath.

However, as a perspective, we can provide a constraint analysis via Dirac and symplectic formalisms, in order to compare the before and after soldering.   The comparison can also be made together with commutative models, namely, what is new in the NC introduction.  Another path is to investigate soldering in the light of the canonical  noncommutativity, where the NC parameter is constant. 

The conversion of NC second-class constraints into first-class ones concerning the soldered actions can reveal interesting properties evolving gauge invariance of NC models.  These ideas are, as a matter of fact, ongoing research that will be published elsewhere.

\section {Acknowledgments} 

\ni V.N. would like to thank Prof. J. A. Helay\"el-Neto for valuable and insightful discussions. E.M.C.A. thanks CNPq (Conselho Nacional de Desenvolvimento Cient\' ifico e Tecnol\'ogico) through Grants No. 301030/2012-0 and No. 442369/2014-0, for partial financial support, CNPq is  a Brazilian scientific research support federal agency, and the hospitality of Theoretical Physics Department at Federal University of Rio de Janeiro (UFRJ), where part of this work was carried out.

\end{document}